\documentclass[12pt,a4paper]{article}
\usepackage[english]{babel}
\title{Noncommutative Harmonic Oscillator at Finite
Temperature: A Path Integral Approach}
\author{
A. Jahan\\Research Institute for Astronomy and Astrophysics of Maragha,\\ P. O. Box: 55134 - 44, IRAN\\jahan@riaam.ac.ir}

\date{}
\begin{document}

\maketitle
\begin{abstract}
We use the path integral approach to a two-dimensional noncommutative
harmonic oscillator to derive the partition function of the system at finite
temperature. It is shown that the result based on the Lagrangian formulation
of the problem, coincides with the Hamiltonian derivation of the partition
function.
\end{abstract}

\section{Introduction}
 amount of challenge devoted to the subject of
noncommutative space-time. Although, the idea of noncommutating space-time coordinates is
an old proposal [1], the recent discoveries in string/M theories were the main source o
Recently, there has been an immense amount of challenge devoted to the subject of
noncommutative space-time. Although, the idea of noncommutating space-time coordinates is
an old proposal [1], the recent discoveries in string/M theories were the main source of
renewed interests in the subject [2]. In particular, much work is dedicated to study the
noncommutative version of the quantum field theory (see for example [3] and references there
in). In noncommutative field theories the space-time coordinates $x^{\mu}$ are replaced by the noncommutating coordinates
operators $\hat{x}^{\mu}$ , satisfying
\begin{equation}\label{eq:1}
[\hat{x}^{\mu},\hat{x}^{\nu}]=i\theta^{\mu\nu}
\end{equation}
with $\theta^{\mu\nu}$ as a real anti-symmetric matrix. Thus because of the uncertainty relation induced
by the relation (1), the short distance scales in the $x^{\mu}$  direction correspond to the large
distance scales in the $x^{\nu}$ direction and vice versa. So there will be a mixing between the
ultraviolet and infrared behaviors of the field theories in noncommutative space-times. Such
a problem, which is called the "UV/IR mixing problem", is characteristic of the models
defined in noncommutative space [4]. In a noncommutative space–time the usual product
between the fields must be replaced by the Weyl–Moyal bracket or star-product defined as
\begin{equation}\label{eq:2}
f(x)\star{g(x)}=\lim_{x\to{y}}\exp(\frac{i}{2}\theta^{\mu\nu}\partial^{x}_\mu\partial^{y}_\nu)f(x)g(y)
\end{equation}
The star-product is the usual starting point in the most studies about the noncommutative
field theories since it encodes the noncommutativity of space–time for the product of the
several fields defined in the same point in the following sense
\begin{equation}\label{eq:3}
\hat{f}_{1}(\hat{x})\cdots\hat{f}_{n}(\hat{x})\to{f_{1}(x)\star}\cdots\star{f_{n}(x)}
\end{equation}
On the other hand it seems natural to explore the noncommutative version of quantum
mechanics since it stands as the one-particle low energy sector of the quantum field theory
[5]. The following set of relations among the coordinates and momenta (setting $\hbar=1$)
{\setlength\arraycolsep{2pt}
\begin{eqnarray}\label{eq:4}
[\hat{x}^{i},\hat{x}^{j}]&=&i\theta^{ij}\\
{}[\hat{x}^{i},\hat{p}^{j}]&=&i\delta^{ij}\nonumber\\
{}[\hat{p}^{i},\hat{p}^{j}]&=&0\nonumber
\end{eqnarray}}
characterize the noncommutative version of the quantum mechanics (Latin indices stand for
the spatial coordinates). An equivalent procedure to implement the star-product is to define
a new set of commutating coordinates $x^{i}$ via [5]
\begin{equation}\label{eq:5}
x^{i}=\hat{x}^{i}+\frac{1}{2}\theta^{ij}p^{j}
\end{equation}
(Through the paper summation is implied over the repeated indices). Therefore for the
interaction potential $\hat{U}(\hat{x}^{i})$ defined in noncommutative space one gets the effective potential
in usual commutative space as
\begin{equation}\label{eq:6}
U(x^{i},\bar{p}^{i})=U(x^{i}-\theta^{ij}p^{j}/2),\quad\bar{p}^{i}=\theta^{ij}p^{j}
\end{equation}
Path integral approach to the noncommutative quantum mechanics has revealed in series of
the recent works [6]. It is demonstrated by the authors that how the noncommutativty of
space could be implemented in Lagrangian formulation of the quantum mechanics. Here we
aim to incorporate the effect of noncommutativty of space on the thermodynamics of a two
dimensional harmonic oscillator by means of the formalism developed in [6]. So after a brief
review of the Lagrangian aspect of noncommutative quantum mechanics we derive the
partition function of the system by identifying the temperature as imaginary time. It is
shown that the result is in accordance with one obtained via the Hamiltonian approach to
the partition function.
\section{Lagrangian Formulation of Noncommutative Quantum Mechanics}
The Hamiltonian governing the dynamics of a harmonic oscillator in noncommutative space is
\begin{equation}\label{eq:7}
\hat{H}=\frac{1}{2m}(\hat{p}^{2}_{1}+\hat{p}^{2}_{2})+\frac{1}{2}m\omega^{2}(\hat{x}^{2}_{1}+\hat{x}^{2}_{2})
\end{equation}
On implementing the transformation (5), one finds the effective Hamiltonian in usual
commutative space as
\begin{equation}\label{eq:8}
H_{\theta}=\frac{\kappa}{2m}(p^{2}_{1}+p^{2}_{2})+\frac{1}{2}m\omega^{2}(x^{2}_{1}+x^{2}_{2})
+\frac{1}{2}m\theta\omega^{2}(x_{1}p_{2}-x_{2}p_{1})
\end{equation}
with $\kappa=1+m^{2}\omega^{2}\theta^{2}/4$ and corresponding energy spectrum as
\begin{equation}
\varepsilon_{n_{+},n_{-}}=\omega[(\sqrt{\kappa}+\sqrt{\kappa-1}\,)n_{+}+(\sqrt{\kappa}-\sqrt{\kappa-1}\,)n_{-}
+\sqrt{\kappa}\,]
\end{equation}
The corresponding Lagrangian can be achieved by means of the Legender transformation $
L_{\theta}(x^{i},\dot{x}^{i})=p^{i}\dot{x}^{i}-H_{\theta}(x^{i},p^{i})$, upon replacing for the momenta from
$\dot{x}^{i}=\partial{H_{\theta}}/\partial{p^{i}}$ Hence the Lagrangian associated with the Hamiltonian (8) will be
\begin{equation}\label{eq:10}
L_{\theta}=\frac{m}{2\kappa}(\dot{x}^{2}_{2}+\dot{x}^{2}_{1})-\frac{m\omega^{2}}{2\kappa}
(x^{2}_{2}+x^{2}_{1})+\frac{\theta{m^{2}}\omega^{2}}{2\kappa}(\dot{x}_{2}x_{1}-\dot{x}_{1}x_{2})
\end{equation}
The above Lagrangian admits the equations of motion as differential equations of rank four.
When the solutions are inserted in (10) and integrated out over the time variable $\tau$ , one
finds for the action
{\setlength\arraycolsep{2pt}
\begin{eqnarray}\label{eq:11}
S_{\theta}(\textbf{x}^{\prime\prime},\textbf{x}^{\prime},\tau)
&=&\frac{m\omega}{2\sqrt{\kappa}\sin(\omega\tau\sqrt\kappa)}\bigg[(\textbf{x}^{\prime\,{2}}+
\textbf{x}^{\prime\prime\,{2}})\cos(\omega\tau\sqrt{\kappa})\\
&-&2(\textbf{x}^{\prime}\cdot{\textbf{x}^{\prime\prime}})
\cos(\omega\tau\sqrt{\kappa-1})\nonumber
+2(\textbf{x}^{\prime}\times{\textbf{x}^{\prime\prime}})_{z}\sin(\omega\tau\sqrt{\kappa-1})\bigg]\nonumber
\end{eqnarray}
}
with boundary condition $\textbf{x}^{\prime\prime}=\textbf{x}(\tau)$ and $\textbf{x}^{\prime}=\textbf{x}(0)$.
Thus one obtains the semiclassical propagator (transition amplitude) as
{\setlength\arraycolsep{2pt}
\begin{eqnarray}\label{eq:12}
K_{\theta}(\textbf{x}^{\prime\prime},\tau;\textbf{x}^{\prime},0)&=&\sqrt{det\bigg(-\frac{\partial^{2}S_{\theta}}
{\partial\textbf{x}^{\prime\prime}\partial\textbf{x}^{\prime}}\bigg)}e^{iS_{\theta}}\\
&=&\frac{m\omega}{2\pi{i}\sqrt{\kappa}|\sin(\omega\tau\sqrt{\kappa})|}e^{iS_{\theta}}\nonumber
\end{eqnarray}}
\section{Finite temperature considerations: Partition function}
The partition function Z($\beta$) plays a vital role in thermodynamical considerations of the
physical systems at finite temperature. It can be defined in terms of the propagator of system
as [7]
\begin{equation}\label{eq:13}
Z(\beta)=TrK(\textbf{x}^{\prime\prime},\beta;\textbf{x}^{\prime},0)
\end{equation}
where the inverse temperature parameter is defined as $\beta=i\tau$. The symbol $Tr$ stands for the
functional trace which for a bi-local function $A(\textbf{x},\textbf{x}^{\prime})$ in \textit{D} dimensions is defined as
\begin{equation}\label{eq:14}
TrA(\textbf{x},\textbf{x}^{\prime})=\int_{-\infty}^{+\infty}d^{D}{x}A(\textbf{x},\textbf{x})
\end{equation}
When the time parameter in (12) is replaced with the inverse temperature parameter
$\beta=i\tau$, the semiclasical propagator (12) modifies to
\begin{equation}\label{eq:14}
K_{\theta}(\textbf{x}^{\prime\prime},\beta;\textbf{x}^{\prime},0)=\frac{m\omega}{2\pi\sqrt{\kappa}\sinh(\omega\beta\sqrt{\kappa})}
e^{-S^{E}_{\theta}}\nonumber
\end{equation}
with
{\setlength\arraycolsep{2pt}
\begin{eqnarray}\label{eq:15}
S^{E}_{\theta}(\textbf{x}^{\prime\prime},\textbf{x}^{\prime},\beta)
&=&\frac{m\omega}{2\sqrt{\kappa}\sinh(\omega\beta\sqrt\kappa)}\bigg[(\textbf{x}^{\prime\,{2}}+
\textbf{x}^{\prime\prime\,{2}})\cosh(\omega\beta\sqrt{\kappa})\\
&-&2(\textbf{x}^{\prime}\cdot{\textbf{x}^{\prime\prime}})
\cosh(\omega\beta\sqrt{\kappa-1})\nonumber
+2(\textbf{x}^{\prime}\times{\textbf{x}^{\prime\prime}})_{z}\sinh(\omega\beta\sqrt{\kappa-1})\bigg]\nonumber
\end{eqnarray}}
where we have invoked the well-known trigonometric identities $\cos{i\alpha}=i\cosh\alpha$ and
$\sinh\alpha=-i\sin{i}\alpha$. Thus from the definition (14) we find for the trace of the term $e^{-S^{E}_{\theta}}$
\begin{equation}\label{eq:21}
Tre^{-S^{E}_{\theta}}=\frac{\pi\sqrt{\kappa}}{m\omega}\frac{\sinh(\omega\beta\sqrt{\kappa})}{\cosh(\omega\beta\sqrt{\kappa})
-\cosh(\omega\beta\sqrt{\kappa-1})}
\end{equation}
Therefore one if left with the partition function as
{\setlength\arraycolsep{2pt}
\begin{eqnarray}\label{eq:16}
Z_{\theta}(\beta)&=&\frac{m\omega}{2\pi\sqrt{\kappa}\sinh(\omega\beta\sqrt{\kappa})}
Tre^{-S^{E}_{\theta}}\\
&=&\frac{1}{2\big[\cosh(\omega\beta\sqrt{\kappa})
-\cosh(\omega\beta\sqrt{\kappa-1})\big]}\nonumber\\
&=&\frac{1}{4\sinh\big[\frac{\omega\beta}{2}(\sqrt{\kappa}+\sqrt{\kappa-1})\big]
\sinh\big[\frac{\omega\beta}{2}(\sqrt{\kappa}-\sqrt{\kappa-1})\big]}\nonumber
\end{eqnarray}}
The above result for the partition function is derived earlier in the context of Hamiltonian
formulation of the problem [8]
{\setlength\arraycolsep{2pt}
\begin{eqnarray}\label{eq:17}
Z_{\theta}(\beta)&=&tr{e}^{-\beta{H_\theta}}
=\sum_{n_{+},n_{-}=0}^{+\infty}e^{-\beta\varepsilon_{n_{+},n_{-}}}\\
&=&\frac{1}{4\sinh\big[\frac{\omega\beta}{2}(\sqrt{\kappa}+\sqrt{\kappa-1})\big]
\sinh\big[\frac{\omega\beta}{2}(\sqrt{\kappa}-\sqrt{\kappa-1})\big]}\nonumber
\end{eqnarray}}
with energy spectrum given by Eq.(9).
The free energy $F_{\theta}(\beta)$ of a system at finite temperature is related to the partition function as
\begin{equation}\label{eq:18}
F_{\theta}(\beta)=-\frac{1}{\beta}\ln{Z_{\theta}(\beta)}
\end{equation}
which at low temperature limit tends to the ground state energy of the system. In particular
for the case of harmonic oscillator we have
\begin{equation}\label{eq:19}
\lim_{\beta{\to}\infty}F_{\theta}(\beta)=\omega\sqrt{\kappa}
\end{equation}
which coincides with the ground state energy of the system, $\varepsilon_{0,0}=\omega\sqrt{\kappa}$ (see Eq. (9)). In the
limit $\theta{\to}0$ , the parameter $\kappa$ tends to unity and one recovers the ground state energy of
usual two-dimensional harmonic oscillator.

\end{document}